\documentstyle[12pt,epsf,aaspp4]{article}

  \def\vol#1  {{{#1}{\rm,}\ }}
  
  \def\etal{et al.\ }

  \newcount\refno
  \newcount\rfno
  \def\eq{$^{\the\refno\ }$\advance\refno by 1}
  \def\ad{\advance\rfno by 1}

  \def\clock{\count0=\time \divide\count0 by 60
  \count1=\count0 \multiply\count1 by -60
 \advance\count1 by \time
           
 \number\count0:\ifnum\count1<10{0\number\count1}\else\number\count1\fi}

\def\Gcm2{\rm G~cm^2}

\def\beq{\begin{equation}}
\def\eeq{\end{equation}}

\def\h2{${\rm\,H_2}$}

\def \date         {\ifcase\month \message{zero} \or
                    January \or February \or March \or
				  April \or May \or June 
                   \or July \or 
                         August \or September \or October
		  \or November \or 
									                      December \fi
                \space\number\day, \number\year}

\begin{document}
\title{A Two-Fluid Thermally-Stable Cooling Flow Model}
\author{Renyue Cen\altaffilmark{1}}
\altaffiltext{1} {Princeton University Observatory,
		  Princeton University, Princeton, NJ 08544;
		  cen@astro.princeton.edu}
%\received{\date}
%\accepted{ }

\begin{abstract}

A new model for cooling flows in X-ray clusters,
capable of naturally explaining salient features observed,
is proposed.
The only requirement is that a significant relativistic component, 
in the form of cosmic rays (CR), 
be present in the intra-cluster medium
and significantly frozen to the thermal gas.
Such an addition qualitatively alters the 
conventional isobaric thermal instability criterion such that
a fluid parcel becomes thermally stable when 
its thermal pressure drops below a threshold fraction of its CR pressure.
Consequently, the lowest possible temperature at
any radius is about one third of the ambient temperature
{\it at that radius}, exactly as observed,
In addition, we suggest that dissipation of internal 
gravity waves, excited by radial oscillatory motions
of inward drifting cooling clouds about their radial equilibrium positions,
may be responsible for heating up cooling gas.
With the ultimate energy source for powering the cooling X-ray luminosity
and heating up cooling gas being gravitational due to inward drifting
cooling clouds as well as the general inward flow,
heating is spatially distributed
and energetically matched with cooling.
One desirable property of this heating mechanism
is that heating energy is strongly centrally concentrated, providing 
the required heating for emission-line nebulae.

\end{abstract}


\section{Introduction}

It is a firmly established observational fact 
that a substantial central region of many X-ray clusters 
of galaxies is cooling in X-rays on time scales
that are shorter than their presumed age,
implying that a large amount of gas would cool to lower temperatures,
if not intervened
(e.g., Fabian 1994; White, Jones, \& Forman 1997; Peres \etal 1998; Allen 2000).

The classic picture was that the gas
would cool isobarically and sink to the center
(Fabian \& Nulsen 1977;
Cowie \& Binney 1977; Fabian 1994).
While simple and elegant, this model is at odds with 
current observational evidence,
with two major contradictions.
First, the amount of cooled gas at the centers of clusters, 
although having been detected
in various forms, including UV emitting gas (e.g., Oegerle \etal 2001),
H$\alpha$ emitting gas (e.g., Heckman \etal 1989; Crawford \etal 1999),
molecular gas (e.g., Jaffe \& Bremer 1997; 
O'Dea \etal 1998; Donahue \etal 2000
Edge 2001; Salome \& Combes 2003; 
Edge \& Frayer 2003)
and star formation (e.g., Fabian \etal 1991;
Hansen, Jorgensen, \& Norgaard-Nielsen 1995; Allen 1995; Smith \etal 1997),
appears to be far short of what
is implied by the cooling rate in X-rays.
This is the ``classic cooling-flow problem" (Peterson \etal 2003).
Second, the supposedly
cooled X-ray gas does not appear to show up 
at temperatures lower than about one third of the maximum temperature
at each radius (e.g., Peterson \etal 2001,2003; %A1835 XMM 10^46 erg/s z=0.25 kT=8.2keV, kT_floor=2.7keV
%Peterson \etal 2003; %spectroscopic constraints on CF models for CLs, much less gas at T<T_x/3
Kaastra \etal 2004);
%which in turn is decreasing moderately towards the center;
i.e., the expected soft X-rays are missing (e.g., Fabian \etal 2002) 
--- the ``soft X-ray cooling-flow problem" (Peterson \etal 2003).

In light of these observational developments various amendments
to the standard cooling flow model with isobaric thermal instability
have been proposed to change the state
of the X-ray cooled gas based on energetic considerations.
There are mostly two schools of ideas, 
either with heating by some sources or 
by cooling in channels other than soft X-rays.
The reader is referred to Mathews \& Brighenti (2003) and 
Peterson \etal (2003) for up-to-date assessments
of various proposed models, confronted 
with observations.
In short, it appears that none of the proposed amendments is fully
satisfactory.

In this {\it Letter} we put forth a new model, based on the assumption that 
a significant fraction of total pressure of the X-ray cluster gas 
is in the form of cosmic rays, 
an unavoidable by-product 
formed at the ubiquitous high velocity shocks 
during the formation of X-ray clusters.
We will show that this model explains 
all major characteristics of observed X-ray cooling flows naturally.

\section{A Two-Fluid Model}
Our purpose is to present a viable physical model
for the cooling flows in X-ray clusters.
Rigorous calculations possibly require detailed simulations
and are not attempted here.

\subsection{Cosmic-Ray Pressure in the Intra-Cluster Medium}
Astrophysical shocks have long been recognized as efficient
sites for CR acceleration (Blandford \& Ostriker 1978).
Clusters of galaxies are full of shocks under the action
of gravity, including accretion shocks (e.g., Bertschinger 1985)
and internal shocks (e.g., Miniati \etal 2000), 
and hence should be excellent production sites of CR.
Miniati \etal (2001), through detailed shock acceleration process of CR,
find that shocks, especially internal high energy (due to high density
and high velocity) shocks within clusters
can convert a large fraction of gravitational 
energy released during the formation of X-ray clusters into CR.
Miniati \etal (2001) show that a few tens of percent of the total
pressure in the intra-cluster medium (ICM) may be in the form of CR,
while Ryu \etal (2003) subsequently show that about $1/3$ of the total
pressure in the ICM may be in CR.
We define $\alpha\equiv P_{CR}/P_t$,
where $P_{CR}$ and $P_t$ are the CR and total pressure, respectively,
and for simplicity $\alpha$ is assumed to be
constant spatially. 
It is assumed that thermal gas is
coupled on the scales concerned to CR by magnetic field.
Significant magnetic field is known to exist
in clusters of galaxies
(e.g., Kim \etal 1990; Taylor \& Perley 1993; Feretti \etal 1995),
although for the purpose of coupling CR to thermal gas a strong
magnetic field is not required.
Observationally, evidence for the existence 
of a significant CR pressure in clusters of galaxies
is also strong (e.g., Blasi 1999). 
In passing, we note that our conclusions will be the same,
if CR pressure is replaced by another form of relativistic component
such as tangled magnetic fields.

\subsection{Dynamics of Cosmic Ray Modulated Cooling Flows}
Consider a somewhat overdense parcel of gas containing a significant fraction 
of CR pressure in X-ray clusters.
Such a parcel would normally tend to fall towards the center, because of 
anti-buoyancy and thermal instability.
We show that, if CR are frozen to the thermal gas in the parcel,
the thermal gas becomes thermally stable below a certain temperature.
Assuming the ratio of specific heats for CR to be $4/3$
and denoting the ratio of the thermal gas pressure to the CR pressure 
in the parcel as $\beta$,
then, the standard (local) isobaric thermal instability criterion (Field 1965; Balbus 1985) can be shown to become
\begin{equation}
\left({\partial\zeta\over\partial T}\right)_\rho 
-\left({3\beta\over 3\beta+ 4}\right) \left({\rho\over T}\right) \left({\partial\zeta\over\partial \rho}\right)_T < 0.
\end{equation}
\noindent
Here $\zeta = A \rho T^{1/2} - \eta(\rho)$
is the net cooling rate per unit mass,
where the first term is the Bremsstrahlung cooling with $A$ being a constant
and the second term $\eta$ is the total heating term, 
which may be a function of many things.
We will consider two simple cases: 1) $\eta$ is constant, meaning that
heating is uniform per unit mass;
2) $\eta\propto \rho^{-1}$, meaning that 
heating is uniform per unit volume.
It seems plausible that these two cases may bracket possible real situations.
The following analysis may be straight-forwardly 
extended to more complex heating functions which may also
depend on gas temperature.
Inserting $\zeta$ for the first case with equal mass heating
into equation (1), 
the thermal instability criterion simplifies to 
\begin{equation}
\beta > {4\over 3}.
\end{equation}
\noindent
For the second case with uniform volume heating 
we make the following assumption in order to 
allow further progress: there is no net cooling at the ambient gas
at each radius to normalize the heating term, i.e.,
$\zeta=A\rho T^{1/2} - A\rho_a^2 T_a^{1/2}\rho^{-1}$,
where $\rho_a$ and $T_a$ are
the ambient gas density and temperature, respectively;
with this form equation (1) becomes 
\begin{equation}
\beta > {4\over 3(1+2({\rho\over \rho_a})^{-2}({T\over T_a})^{-1/2})}.
\end{equation}
\noindent
%Note that a pure thermal gas parcel with $\beta=\infty$
%always satisfies the above inequality and is thus 
%isobarically thermally unstable,
%as is well known.
In both cases we see that the conventional picture
of thermal instability in X-ray cluster cooling flows
(Mathews \& Bregman 1978;
Fabian, Nulsen, \& Canizares 1984;
Nulsen 1986) applies only in a limited sense, i.e.,
within a finite temperature range. 
But, 
in the presence of CR pressure,
ultimately, 
there will be no runaway thermal instability and
a temperature floor is put in place
and further cooling of clouds 
relative to the ambient gas ceases.

In the finite range of temperature where thermal instability
does apply, the denser gas parcel cools, gets compressed and tends to fall inward.
Since CR virtually do not cool, the CR pressure increases
relatively during this infall process.
It would finally reach a quasi-equilibrium 
position at a smaller radius in the comoving flow,
with the exact location
depending on the density and temperature 
run in the cooling flow region. 
Two conditions are met there.
First, the gas parcel is in dynamical equilibrium at that radius 
which implies $\rho_{p} = \rho_a$,
where $\rho_p$ and $\rho_a$ are the gas density in the parcel and
ambient gas density, respectively.
Second, the parcel needs to be in pressure equilibrium with the ambient gas,
requiring $p_{p}=p_a$, 
where $p_{p}=(1+1/\beta) \rho_{p} k T_{p}/m_p$
and $p_a={1\over 1-\alpha} \rho_a k T_a/m_p$
are the total pressure in the parcel and in the ambient gas, respectively,
$k$ is the Boltzmann's constant,
$m_p$ is the proton mass.
Combining these relations gives
\begin{equation}
T_{p} = {\beta \over (1-\alpha)(1+\beta)} T_a.
\end{equation}
\noindent
%Note that in the limit $\alpha=0$,
%the only solution observing both dynamical and pressure equilibria
%is $T_p=T_a$.
%But since the gas is thermally unstable in the case of $\alpha=0$,
%this trivial solution tells that there is no equilibrium position for a cooling
%gas parcel.
Thus, a floor value for $\beta$ for thermal instability translates to a floor
in temperature.
We find that for the case with constant heating per unit mass
the lowest possible temperature for a cooling gas parcel is 
\begin{equation}
T_{p} = {4 \over 7} {1\over (1-\alpha)} T_a,
\end{equation}
\noindent
whereas for the case with constant heating per unit volume 
(and utilizing dynamical equilibrium condition $\rho=\rho_a$),
we obtain 
\begin{equation}
T_{p} = {18 + {28\over 1-\alpha} - 6\sqrt{9+{28\over 1-\alpha}}\over 49} T_a.
\end{equation}
\noindent
Miniati \etal (2001) show 
that $\alpha$ is about a few tens of percent,
while Ryu \etal (2003) show $\alpha\sim 1/3$.
Thus, using their predicted range
$\alpha=10-30\%$ we obtain the floor temperature
\begin{equation}
T_{p} = (0.6-0.8) T_a,
\end{equation}
\noindent
if heating is uniform in mass,
and 
\begin{equation}
T_{p} = (0.2-0.3) T_a,
\end{equation}
\noindent
if heating is uniform in volume.
This indicates that the temperature floor may somewhat vary,
depending on detailed dependence of heating on density.
But the essential point is that there will always
be an absolute floor at about one third of the ambient
temperature at any radius.
Thus, it is borne out in our model, 
{\it without any fine tuning},
that the coldest gas at each radius that coexists
with the hotter ambient gas may have a temperature of
about one third of that of the ambient gas {\it at that radius},
explaining the seemingly puzzling
observations (Peterson \etal 2003; Kaastra \etal 2004).

Equation (8) only demands
that the ratio of the coldest temperature to ambient temperature
at any radius maintain a certain ratio.
However, it does not prevent overall ambient temperature
$T_a$ from decreasing or increasing.
In the cooling flows $T_a$ would decrease 
in the absence of a balanced heating source.
In the next section we suggest a heating
mechanism that is physically based and has
the desired properties.

\subsection{Heating by Dissipation of Internal Gravity Waves}

The rising gas density and dropping gas temperature towards
the center of cooling flows 
produce a subadiabatic run and therefore gas is 
convectively stable under the Schwarschild criterion for {\it adiabatic} gas.
However, for gas parcels with temperature ranging from $\sim 1/3T_a$ to $T_a$, 
we have a thermally unstable gas,
which is also convectively unstable even for a subadiabatic run (Defouw 1970).
But for a strongly subadiabatic run, as in the cooling flow regions
where cooling time may be longer than the inverse of oscillation frequency,
it is expected that the thermal-convective instability
actually becomes overstable, first suggested analytically (Defouw 1970)
and subsequently confirmed by numerical simulations (Malagoli, Rosner, \& Fryxell 1990),
manifested by the radial oscillation of a cloud about its equilibrium position 
with an exponentially growing amplitude,
instead of monotonic inward motion.
In our case, since clouds can not cool down below a certain temperature
(equations 7,8), overstability is unavoidable.
At each radius a cloud would oscillate about its equilibrium radial 
position with its local
buoyant oscillation (Brunt-V\"ais\"al\"a) 
frequency, 
$\omega_{BV}^2=\Omega_c^2({3\over 5}{d\ln T\over d\ln r}-{2\over 5}{d\ln \rho\over d\ln r})$,
where $\Omega_c$ is orbital frequency.
We suggest that such radial oscillations would provide a natural
mechanism to excite acoustic and gravity waves.
Heating of cooling flow regions by 
acoustic waves has been considered before 
(Pringle 1989)
and will not be considered further here.
%partly because it is not clear how much acoustic wave energy
%can be channeled to heat the cooling regions.
We will instead focus on the internal gravity waves.

In cooling flow regions,
$T$ and $\rho$ each may be assumed to bear a power-law form,
$T\propto r^{\gamma_T}$ and $\rho\propto r^{\gamma_\rho}$,
and if the dark matter density 
is assumed to be $\rho_{DM}\propto r^{\gamma_{DM}}$,
then the Brunt-V\"ais\"al\"a frequency
$\omega_{BV}^2 \propto r^{\gamma_{DM}} (\gamma_T-{2\over 3}\gamma_\rho)$
with a positive front coefficient.
For normal X-ray clusters, 
$\gamma_{DM}<0$,
$\gamma_{T}>0$,
$\gamma_{\rho}<0$,
thus, $\omega_{BV}$ is positive
and increases monotonically
towards the center of the cluster.
Gravity waves with frequency $\omega > \omega_{BV}$ are evanescent.
Therefore, in our case where oscillatory motions of cooling clouds
are responsible for exciting the internal gravity waves,
the fact that the buoyant oscillation frequency 
increases towards the center of the cluster
suggests that any internal gravity waves 
generated by radially oscillating clouds
at a certain radius will be trapped to the region interior to that radius.
These trapped
internal gravity waves will eventually dissipate to heat the gas.
When considering wave excitement by orbiting galaxies (Miller 1986),
Balbus \& Soker (1990) and 
Lufkin, Balbus, \& Hawley (1995) pointed out earlier  
that the central regions of cooling flow clusters 
are capable of trapping finite-amplitude resonant internal gravity waves,
when the orbiting frequency of galaxies falls below the 
Brunt-V\"ais\"al\"a frequency at a certain radius.
However, Lufkin \etal (1995) concluded that internal gravity waves
excited by orbiting galaxies 
may not be able to provide enough energy to balance the cooling
in the cooling flow region.

The energy source for powering the oscillations of clouds
hence the excited internal gravity waves
in our case is gravitational potential energy of cooling gas. 
Tapping into the gravitational energy to heat up cooling flows
is not a new idea and its desirable property of being able
to provide overall
balanced heating has been recognized (e.g., Fabian 2003).
Here, gravitational energy released by a cloud 
originating at some radius is telescopically 
deposited within some smaller radius 
at which it finds its quasi-equilibrium position in the comoving flow.
Heating is hence ultimately due to gravitational energy released
as a direct consequence of cooling and inward drift of clouds
as well as the slower and subsonic inward motion of the general flow.
Since heating is in essence ``reactive" to cooling, 
cooling should be balanced by heating over time,
although some spatial and temporal fluctuations may be expected.
An energetic advantage of heating by the trapped 
gravity waves is that it is economical without significant {\it waste}
because of trapping.

Another interesting property of this telescopic heating by gravity waves
is that all excited waves at radius larger than $r$ would contribute
to heating of gas at $r$.
A somewhat more quantitative and relevant illustration 
of heating by internal gravity waves
may be made as follows.
Assuming mass dropout rate per unit radius is constant,
as observed, i.e., $d\dot M/dr = C$, in the cooling region $r=0-R$,
assuming that the temperature scales with radius as
$T(r)=T_0(r/R)^{\gamma_T}$ and temperature traces local
gravitational potential,
$kT(r)/m_p=GM(<r)/2r$,
and gas density goes as $\rho(r)=\rho_0(r/R)^{\gamma_\rho}$,
assuming that interval gravity wave energy is deposited
uniformly in radius interior to its trapping radius,
assuming that each cloud's gravitational potential energy 
going from its initial position ($r_i$) to the final position ($r_f$),
$\Delta\psi\equiv \int_{r_i}^{r_f} g(r) dr$, is released interior
to its equilibrium position,
assuming that CR pressure evolves adiabatically from $r_i$ to $r_f$,
and assuming that all gas clouds that move inward
cool to the allowed floor temperature,
then a simple calculation yields the gravity wave energy deposition
rate interior to a radius $r$ to be equal to
\begin{eqnarray*}
\dot E(<r)&=&\int_{r_i}^R {r\over x} C {\Delta\psi\over m_p}dx 
+\int_0^{r_i} C {\Delta\psi\over m_p}dx \\
&=& {2\dot M(<r) kT(r) \over m_p \gamma_T^2}
\left[\left({1+\beta\over 1+\alpha^{-1}}\right)^{3\gamma_T/(\gamma_{\rho}-3\gamma_T)} -1\right] \\
&& \left[\left({R\over r}\right)^{\gamma_T} + 
\left({1+\beta\over 1+\alpha^{-1}}\right)^{3\gamma_T/(\gamma_{\rho}-3\gamma_T)} 
[\left({1\over 1+\gamma_T}\right)\left({1+\beta\over 1+\alpha^{-1}}\right)^{3/(\gamma_{\rho}-3\gamma_T)}
-1]\right]
\end{eqnarray*}
\noindent
where $r_i$ is the initial position of clouds that
find their equilibrium position at $r$.
The normally implied heating rate, assumed to be equal to 
the mass dropout rate (assuming heating balances cooling), would be
$\dot E(<r)|_{local, implied}=\int_0^{r_i} {3CkT(x)\over 2m_p}dx 
={3\dot M(<r) kT(r)\over 2m_p (\gamma_T+1)}$.
For $\alpha=0.3$ and $\beta=0.3$, which give $T_p=0.33T_a$,
and using $\gamma_T=0.75$ and $\gamma_\rho=-0.75$, consistent with
some observed X-ray cluster cooling flows (e.g., Kaastra \etal 2004),
we evaluate the ratio $\dot E(<r)/\dot E(<r)|_{local,implied}=48$ for $R/r=10$,
which is comparable to the energy dropout rate at $R$.
Therefore, it is seen that the telescopic heating
results in a progressively
stronger heating rate at small radii
than would be implied by its local observed mass dropout rate.
This unique property may help explain the large inferred/required
heating rate of $H_\alpha$ emission-line nebulae (Heckman \etal 1989).
But we caution this simple calculation serves only
an illustrative purpose.

A desirable property of this heating mechanism by
gravity waves is self-regulation, as insightfully pointed
out by Balbus \& Soker (1990).
If, say, the central region is overheated such that
the central temperature becomes very high and possibly 
the central temperature inversion is removed,
that would then cause 
the Brunt-V\"ais\"al\"a frequency to drop
towards that central region.
Consequently,
inward propagating gravity waves generated outside
would get reflected at the surface enclosing
the overheated region which marks the peak 
of the Brunt-V\"ais\"al\"a frequency.
Thus, the large amount of heating provided by
waves generated at large $r$ which caused
the overheating in the central region 
is no longer able to deliver energy there;
the heating rate of the overheated region would dramatically
decrease to be unable to balance the cooling
and the temperature would drop,
until a rising Brunt-V\"ais\"al\"a frequency towards
the center is achieved.

Finally, it might be expected that trapped 
quasi-spherical gravity waves perhaps eventually lead to quasi-spherical
dissipation patterns in the cooling regions with the patterns
likely being more visible in the central regions due to
more intense cooling/heating, as quantified above.
This feature might have already been seen
in the Perseus cluster from
recent Chandra observations by Fabian \etal (2003).

\section{Discussion}

The critical assumption made 
is that CR are frozen to thermal gas 
on the cooling time scale.
The diffusion time scale of CR is highly uncertain in the absence
of reliable knowledge of the topology of magnetic field and turbulence.
Nonetheless, it may still be useful to have some estimates, but bearing 
in mind that these are just estimates with large uncertainties.
For pitch-angle scattering diffusion for Alfv\'en turbulence
one finds the diffusion coefficient along the magnetic 
field lines
(e.g., by integrating equation 15a of Dermer, Miller \& Li 1996) to be 
$\kappa_{||}={8v\over \pi}{1\over (q-1)(2-q)(4-q)}k_{min}^{-1}\xi_i^{-1} (r_Lk_{min})^{2-q}p^{2-q}$,
where $v$ is CR velocity, $k_{min}$ the minimum wavenumber of turbulence,
$r_L$ the non-relativistic Larmor radius of CR,
$\xi$ the normalized energy density in either shear Alfv\'en 
or fast-mode waves,
$p$ the dimensionless CR momentum $v/c\sqrt{1-v^2/c^2}$ 
and 
$-q$ the power spectrum index of Alfv\'en turbulence.
The actual diffusion coefficient
depends quite sensitively on 
the power spectrum index of Alfv\'en turbulence, $q$.
For $q=(5/3, 3/2, 4/3)$ we find, respectively,
$\kappa_{||}=1.3\times 10^{29} ({v\over c})({B\over 1\mu G})^{-1/3}({\lambda_{max}\over 1kpc})^{2/3}
\xi_i^{-1}p^{1/3}$~cm$^{2}$s$^{-1}$,
$\kappa_{||}=4.7\times 10^{27} ({v\over c})({B\over 1\mu G})^{-1/2}({\lambda_{max}\over 1kpc})^{1/2}
\xi_i^{-1}p^{1/2}$~cm$^{2}$s$^{-1}$,
and
$\kappa_{||}=7.2\times 10^{25} ({v\over c})({B\over 1\mu G})^{-2/3}({\lambda_{max}\over 1kpc})^{1/3}
\xi_i^{-1}p^{2/3}$~cm$^{2}$s$^{-1}$,
where $\lambda_{max}$ is the maximum wavelength of turbulence.
CR diffusion across magnetic field lines is generally
thought to be much smaller and neglected here.
Thus, during a cooling time of order $10^8$~yr,
CR may diffuse by a distance of order $10$kpc, $3$kpc and $0.3$kpc,
for the three cases.
We see that the assumption of CR being frozen with thermal gas 
may be valid for parcels larger than from $\sim 0.3$kpc to $10$kpc,
but it is entirely possible that the diffusion scale
could be significantly smaller than that range.
This issue clearly deserves further attention, 
most likely to be resolved only with better simulations
based on first principles.

The combined thermal-dynamical process is influenced by four 
time scales: cooling time, the oscillation period (the inverse
of the Brunt-V\"ais\"al\"a frequency), CR diffusion time 
and cloud sound crossing time.
The implicit requirement for isobaric evolution
is that cooling time is longer than sound crossing time
of a gas parcel.
For a parcel of size of $1$kpc, the sound crossing time
for a sound speed of $300$km/s is $3\times 10^6$yr,
whereas the cooling time is $\ge 10^8-10^9$yr.
Thus, for parcels of interest here, it appears that
isobaric condition is met.
The oscillation period is in the range
of $10^7-10^8$yrs in the $r=10-100$kpc range.
Our derivation assumes conservatively that the parcel of gas 
is able to cool rapidly enough to reach its possible
floor temperature before it reaches its radial
equilibrium position. 
For a cloud initially only slightly overdense the ratio
of the final equilibrium radius to its initial radius
is $0.39$ for $\alpha=0.3$ and $\beta=0.3$,
if cooling is rapid enough.
The time of the travel to the final equilibrium position 
should be comparable to or longer than the oscillation period or 
the cooling time, whichever is longer.
The possibility that cooling time is longer than the 
oscillation time indicates that the cloud
can only fall on the cooling time, partially supported
by buoyant force on its way in,
and that the inward motion may not be monotonic.
It may be argued that ultimately CR would diffuse out,
since diffusion time, however long, is likely to be
finite.
Weak diffusion of CR may in fact be necessary
to ultimately enable gas to move inward.
More detailed calculations possibly require joint 
considerations of all the processes involved
most likely in a simulation setting.

Invariably, CR diffusion time must have a certain distribution,
depending probably primarily on the topology of local magnetic field
within a fluid parcel.
For example, a fluid parcel may contain a set of relatively open magnetic
field lines and its CR would be more difficult to trap.
Such a fluid parcel would then continue to fall inward
further than others and would cool to a significantly
lower temperature, in a fashion customary in the 
usual thermal instability picture.
It is instructive to consider a somewhat more general form
of cooling-heating function
$\zeta = A \rho T^{\phi} - \eta$, 
and for illustrative purpose we assume equal mass heating (i.e., $\eta$
is constant; using equal volume heating gives qualitatively 
similar conclusion), for which 
the thermal instability criterion becomes
\begin{equation}
\beta > {3\phi\over 4-4\phi}.
\end{equation}
\noindent
Clearly, for negative $\phi$,
equation (9) is always satisfied hence
the gas always thermally unstable.
One way to understand this is that,
with negative $\phi$,
even isochoric cooling is thermally unstable.
Therefore, gas with temperature less than about
$0.1-1.0$keV, depending on the metallicity
of the gas (e.g., Sutherland \& Dopita 1993),
where Bremsstrahlung cooling no longer dominates
and $\phi$ negative,
is thermally unstable, regardless
of the CR pressure.
Thus, a dense fluid parcel with ``leaky"
CR may be thermally unstable on its way inward.
Once its temperature has dropped below about 
$0.1-1.0$keV, the runaway thermal instability sets in
and its temperature may continue to drop rapidly
until it reaches about $10^4$K, where $\phi$ becomes
highly positive and the gas would again become thermally stable.
These rapidly cooling clouds may end up near the center
of a cooling flow and may be identified with 
emission-line nebulae (Heckman \etal 1989).
Since gas parcels with higher metallicity would transit to 
runaway thermal instability
at a higher temperature in the $0.1-1.0$keV range,
one signature might be that 
most emission-line nebulae may be biased to have 
higher than average metallicity.
Under the same arguments, metals may be ``transported"
in this fashion to the centers of cooling regions,
possibly producing patches of relatively metal rich gas
and unevenness in the metallicity distribution near
the centers of cooling flow regions as well as creating
an overall negative metallicity gradient.
The expected CR pressure within cold clouds
may easily maintain a sound speed of $100-200$km/s
to explain the observed line width of the nebulae.
%Moreover, the wanted shock heating of nebulae clouds
%(Johnstone 1988; Heckman \etal 1989) is expected.

We have neglected thermal conduction in our treatment,
which may be justified given the requirement of a 
significant magnetic field to confine CR.
The actual distribution of gas mass in the 
range $\sim 0.3T_a$ to $T_a$ at each radius,
where thermal instability plays,
would depend on many factors including the density 
fluctuation spectrum and heating function. 
We will not attempt to make further calculation of it here.

We suggest that the floor temperature at
each radius derived in \S 2.2
is not expected to be visibly correlated with other cluster 
properties, since it is solely determined by
shock properties, which are not clearly related to any other quantity.
It is important to remember 
that the floor temperature should also exist  in non-cooling clusters,
such as Coma, as observed (Kaastra \etal 2004),
since a significant CR pressure should exist in all clusters,
if produced by cluster shocks.

\section{Conclusions}

We put forth a new model for cooling flows
in X-ray clusters of galaxies.
The model is based on the adoption 
of the view that a significant fraction
of the ICM pressure is in cosmic rays, 
as simulations have shown 
(Miniati \etal 2001; Ryu \etal 2003)
and observations have indicated (e.g., Blasi 1999).
We show that 
a significant CR pressure changes qualitatively
the nature of the thermal evolution of the cooling gas.
Specifically, the conventional 
isobaric thermal instability condition
is only met for a finite range of gas temperature,
while a fluid parcel becomes thermally stable when 
its thermal pressure drops below a certain fraction of its CR pressure.
%depending on the details of the heating function.
As a result, the lowest possible temperature at
any radius for bulk of the gas 
is close to one third of the ambient temperature.
This explains naturally
the observed puzzling fact (Peterson \etal 2003; Kaastra \etal 2004),
{\it without the need to make any fine tuning}.

Additionally, we suggest that dissipation of internal 
gravity waves, excited by radial oscillatory motions
of inward drifting cooling clouds,
may be responsible for the overall heating to balance cooling.
The gravitational energy of inward moving clouds
is converted to the form of kinetic energy of the clouds,
which then oscillate and transfer their kinetic energy to gravity wave energy,
which in turn is trapped interior to the location
of each wave generator.
Therefore, the ultimate energy source for powering the cooling X-ray luminosity
and heating up cooling gas is gravitational.
The proposed heating process has several desirable properties.
First, it is relatively widespread spatially.
Second, heating is spatially distributed
and globally energetically matched with cooling.
Finally, heating is strongly peaked at the center,
apparently capable of providing adequate heating for the observed
emission-line nebulae.
Possible signatures of such heating include occasional quasi-spherical 
density enhancements, mostly likely seen 
in the inner cooling flow regions due to an expected
rapid drop-off of enhancement with increasing radius.

%\acknowledgments
This research is supported in part by grants AST-0206299 and NAG5-13381.  
I thank Hui Li and Jim Stone for useful discussion.

\end{document}